\documentclass[onecolumn,floatfix,aps,nofootinbib,showkeys,10pt]{revtex4-2}

\usepackage[dvips]{epsfig}
\usepackage[english]{babel}
\usepackage{bbm}
\usepackage{verbatim}
\usepackage{array}
\usepackage{float}
\usepackage{bm} 
\usepackage{amsmath}
\usepackage{amsfonts}
\usepackage{amssymb}
\usepackage{hyperref}
\usepackage[dvipsnames]{xcolor}
\hypersetup{colorlinks=true,linkbordercolor=Red,linkcolor=Red, citecolor=Red}
\usepackage{indentfirst} %primeiro paragrafo com espaço
\usepackage{float}

\usepackage{bbding}
\usepackage{booktabs}
\usepackage{adjustbox}

\def\p{\mbox{\boldmath$\displaystyle\boldsymbol{p}$}}

\usepackage{bbold}
\def\I{\openone}
\def\openone{\mathbb I}

\begin{document}

\title{A note on hidden classes in spinor classification}

\author{R. J. Bueno Rogerio}\email{rodolforogerio@gmail.com}
\affiliation{Centro Universit\'ario UNIFAAT, Atibaia-SP, 12954-070, Brazil.}

\author{R.~T.~Cavalcanti}\email{rogerio.cavalcanti@ime.uerj.br}
\affiliation{Instituto de Matem\'atica e Estat\'istica, Universidade do Estado do Rio de Janeiro (UERJ), Rio de Janeiro-RJ,  20550-900, Brazil.}
\affiliation{Departamento de F\'isica, Universidade Estadual Paulista (UNESP), Guaratinguet\'a-SP, 12516-410, Brazil.}

\author{C.~H.~Coronado Villalobos}
\email{ccoronado@utp.edu.pe}
\affiliation{Universidad Tecnol\'ogica del Per\'u (UTP), Lima-Per\'u.}

\author{J. M. Hoff da Silva} \email{julio.hoff@unesp.br}
\affiliation{Departamento de F\'isica, Universidade Estadual Paulista (UNESP), Guaratinguet\'a-SP, 12516-410, Brazil.}


\date{\today}

\begin{abstract}
The Lounesto classification is a well-established scheme for categorizing spinors based on their physical content, determined by their associated bilinear forms. It consists of six disjoint classes encompassing the known spinors within the context of the standard model of high-energy physics. However, advancements in theories beyond the standard model have opened the door to potential new spinorial adjoint structures, leading to new unforeseen classes. These developments indicate the potential for extending the standard Lounesto classification. This paper explores all possible subclasses that could extend the Lounesto scheme. We highlight the most relevant subclasses by imposing constraints on their corresponding dual structures, thus broadening our understanding of spinors and their applications in theoretical physics.
%\begin{center}
%\vspace{0.25cm}
%\textit{In memory of Dharam Vir Ahluwalia, whose invaluable contributions to the field continue to inspire and guide our endeavors.}
%\end{center}
\end{abstract}

%\pacs{}
%\keywords{}

\maketitle

\section{Introduction}\label{intro}

Spinors and fermionic particles modeled by spinors are fundamental building blocks in describing high-energy physics. These objects are essential in understanding the dynamical laws associated with particle interactions and the kinematic laws associated with space-time symmetries. However, beyond their theoretical and phenomenological importance, spinors are also well-defined mathematical objects with a rich and fascinating structure. Although Dirac's monumental work has been central, guiding much of our understanding of spinors, it is important to recognize that the spinor landscape is broader than the standard Dirac one. For instance, investigations carried out in the last decade have delved into fermions identified as having mass dimension one \cite{Ahluwalia:2016rwl, Lee:2015sqj, Lee:2015jpa, Lee:2014opa, Villalobos:2015xca}. This exploration is driven by the aim of constructing a potential quantum field theory that might explain dark matter \cite{Ahluwalia:2004ab,Ahluwalia:2009rh, Agarwal:2014oaa}.

The physics carried by spinors is closely related to an often overlooked fundamental aspect: individual spinors cannot be observed directly. Their composite nature, through the bilinear covariants, transforms them into physically observable objects. Standard protocols involve determining the so-called Dirac dual of a spinor and then defining the associated bilinear covariants. This common understanding of spinor fields can be found in textbooks on quantum field theory. Moreover, from the mathematical side, a remarkable theorem called the inversion theorem, proposed by Takahashi, provides an illuminating perspective on spinors and bilinears \cite{Takahashi:1982bb}. It states that, under certain conditions, one can derive spinors from the bilinears they generate. Using this theorem and based on the behavior of their bilinear covariants \cite{lounestolivro}, Lounesto introduced a classification for spinors into six disjoint classes. Their main representatives are the Dirac spinors (classes 1, 2, and 3), flag-dipole spinors (class 4), Majorana or flag-pole spinors (class 5), and Weyl or dipole spinors (class 6). The Fierz-Pauli-Kofink (FPK) identities govern the interplay between different bilinear configurations and potential spinor variants \cite{Crawford:1985qg,Takahashi:1982bb}, which give quadratic relations that the bilinear covariants must obey. The possibility of only six classes is strongly tied to the fact that Lounesto used the dual structure of Dirac to classify spinors. While the traditional approach anchored on the Dirac dual structure has been fundamental in modern high-energy physics, the emerging beyond the standard model theories beckons a reconsideration of this foundation. The potential inclusion of new dual structures might not just be a mathematical exercise \cite{HoffdaSilva:2019ykt, HoffdaSilva:2022ixq}, but also a path leading to deeper insights into relevant theoretical extensions.

In establishing a spinor dual, it is relevant to envisage some peculiarities calling attention to possible generalizations \cite{Ahluwalia:2019etz}. For the argument, starting from $\tilde{\psi}=\psi^\dagger\eta$ as an (almost) standard dual candidate, the requirement that the norm is boosted invariant leads to\footnote{Up to additional freedom, particularly relevant for discussions including physics with Lorentz subgroups invariance.} $\{K_i,\eta\}=0$ (for $i=x,y,z$), where $\vec{K}$ is the boost generator. Such a constraint, along with the imposition of $\tilde{\psi}\psi \in \mathbb{R}$ (i. e., real norm), reduces $\eta$ to
\begin{equation}
\eta=\left(\begin{array}{cccc}
\mathbb{0}_{2\times 2} & n\mathbb{1}_{2\times 2}  \\
m\mathbb{1}_{2\times 2} & \mathbb{0}_{2\times 2}
\end{array}\right),
\end{equation} where $n$ and $m$ are real numbers. The norm invariance under rotations requirement does not furnish additional constraints to $\eta$ than the invariance under boosts. Then, further imposing parity invariance to the norm, one arrives at $n=m$ leading to $\eta\sim\gamma^0$, and the Dirac standard dual is reached $\tilde{\psi}=\bar{\psi}$, after a simple scaling. The point to be stressed here is that from the purely algebraic point of view, there are possibilities for the dual, other than the standard Dirac one. The price to pay is sometimes a lack of Lorentz invariance \cite{kos}, or loss of a discrete symmetry invariance (as mentioned above), and finally, the possibility of non-hermitian (eventually pseudo-hermitian \cite{cydhar}) terms in a lagrangian is also in order \cite{nonher}. These departures from the standard case are very often, and quite reasonably, avoided. Nevertheless, they can also serve as first principle sources of theories beyond the standard model since their impacts may be accommodated into a broader physical scope, with manageable and delimited drawbacks. Whatever physical scenario results from these models, if the departure point comes from spinor duals, the formulation's algebraic aspects remain important compasses. The results presented in this paper follow this reasoning, retaining $\gamma_0$ as the 'metric' and opening the possibility for discrete symmetry operators composing the dual. This line of investigation is a continuation of a program started in Ref. \cite{Cavalcanti:2020obq} and, as we shall see, presents the algebraic possibility of hidden subclasses. We shall start our analysis in the classical realm, understanding this as an essential starting point in a robust approach to fermionic quantum field expansion coefficients.    

In the exploration of different duals, some formal aspects are now well established \cite{HoffdaSilva:2019ykt, Cavalcanti:2020obq}. In particular, beyond the Fierz-Pauli-Kofink (FPK) identities, a dual given by $\stackrel{\neg}{\psi}=[\Delta\psi]^{\dag}\gamma_0$ is such that the dual operator $\Delta$ must obey the algebraic constrain \cite{HoffdaSilva:2019ykt}
\begin{align}\label{algebraic_constrain}
\gamma_0\Delta^\dagger\gamma_0 = \Delta.
\end{align}
Equivalently, taking $\Delta$ as a general complex matrix, the above algebraic constrain makes $\Delta$ to have the form of a block matrix, as represented by
\begin{equation}
\Delta=\left(\begin{array}{cccc}
A & B  \\
C & A^\dagger
\end{array}\right),\; \text{with}\; B^\dagger=B \;\;\text{and}\;\;  C^\dagger=C.
\end{equation}
As a consequence, the possible duals are restricted to those obeying both the FPK identities and the algebraic constraint. In the present paper, we are particularly interested in using this fact to find a finer version of the standard Lounesto classification and investigate new subclasses made possible by redefining the Dirac dual structure. We emphasize that from now on, we will be interested in new mathematical possibilities, sometimes foregoing the standard physical content carried by bilinear forms.

In an attempt to further explore/evince the new physics the investigation of duals may point out, we also refer to \cite{alexis} (see also \cite{ahluwalia2022critical}). In \cite{alexis}, the authors analyse a la Weinberg without considering subtleties on degeneracy beyond the spin to be carried by one particle states of the field investigated. Once all the degrees of freedom are taken into account, no violation of constraints coming from the constructive Weinberg's formalism is presented (\cite{Ahluwalia:2023slc,Ahluwalia:2022yvk}). Such a discussion is prominent since, as we shall see, it seems to be a characteristic of dealing with new duals, the appearance of complementary degeneracies beyond the spin, so that the spin sums are well-behaved.

The paper is organized as follows: In the following section, we shall introduce the basics of the space-time Clifford algebra, delve into the Lounesto classification, and key aspects of the FPK identities. Then, in Sec. \ref{new_subclasses}, we exploit the possible subclasses of the standard Lounesto classification, establishing which ones are allowed by the FPK identities and the algebraic constraint of the general spinor dual \cite{HoffdaSilva:2019ykt,Cavalcanti:2020obq}. As mentioned, we are particularly interested in redefining the duals by applying combinations of the discrete symmetry operators, $\mathcal{C}, \mathcal{P}, \mathcal{T}$, and elements of the space-time algebra. In Sec. IV, we explore different dual structures, giving a complete account of the usage of new duals and the resultant physical scenario. In Sec. \ref{conclusion}, we scrutinize the consequences of the new subclasses built upon the new duals.

\section{Basic conceptions of the usual spinor classification}\label{basic_classification}

Let $M$ represent the Minkowski space-time. Spinor fields are objects of the spinor bundle $\mathbf{P}_{\mathrm{Spin}_{1,3}^{e}}(M)\times_{\tau }\mathbb{C}^{4}$ associated with $M$, carrying $\tau={\left(1/2, 0\right)}\oplus{\left(0, 1/2\right)}$ representations of the spin group \cite{Mosna:2002fr, Rodrigues:2002sg, Vaz:2016qyw}. It is well-known that their observable properties are encoded in the bilinear covariants, known as sections of the exterior bundle $\Omega(M)$ \cite{lounestolivro, Crawford:1985qg}. These symmetry-preserving bilinear covariants arise from the multi-vector structure of the space-time Clifford algebra. In the standard Dirac theory, they are given by
\begin{eqnarray}
\label{covariantes}
\sigma=&\psi^{\dag}\gamma_{0}\psi \in\Omega^0(M),  \\
 \omega=&-\psi^{\dag}\gamma_{0}\gamma_{0123}\psi \in\Omega^4(M),\\ 
 \mathbf{J}=&\psi^{\dag}\mathrm{\gamma_{0}}\gamma_{\mu}\psi\;\gamma^{\mu} \in\Omega^1(M),\\
 \mathbf{K}=&\psi^{\dag}\mathrm{\gamma_{0}}\gamma_{\mu}\textit{i}\mathrm{\gamma_{0123}}\psi\;\gamma^{\mu} \in\Omega^3(M),\\
\mathbf{S}=&\psi^{\dag}\mathrm{\gamma_{0}}\textit{i}\gamma_{\mu\nu}\psi\gamma^{\mu}\wedge\gamma^{\nu} \in\Omega^2(M).
 \end{eqnarray}
Bearing in mind that the Lounesto classification is built upon the usual Dirac adjoint $\bar{\psi} = \psi^{\dag}\gamma_0$. Other extensions of the spinor classification can be found in \cite{beyondlounesto,rodolfotaka}.

Let ${\cal{M}}(4,\mathbb{C})$ represent the algebra of $4\times 4$ matrices with complex entries. We consider the set ${\mathbbm{1},\gamma_{I}}$, where $I\in{\mu, \mu\nu, \mu\nu\rho, {5}}$ is a composite index. This set serves as a basis for ${\cal{M}}(4,\mathbb{C})$, satisfying the Clifford algebra relation $\lbrace\gamma_{\mu }\gamma_{\nu}\rbrace=2\eta_{\mu \nu }\mathbbm{1}$, where $\eta_{\mu\nu}$ represents the Minkowski metric in the mostly negative signature.
According to the usual conventions, the homogeneous elements of the space-time Clifford algebra are chosen in accordance with the above construction \cite{Crawford:1985qg}
\begin{equation}\label{cliffordbasis}
\Gamma = \lbrace \mathbbm{1}, \gamma_{\mu}, \gamma_{0123}, \gamma_{\mu}\gamma_{0123}, \gamma_{\mu}\gamma_{\nu}\rbrace. 
\end{equation}

The well-established classes of spinors include the Weyl, Dirac, and Majorana spinors; however, as mentioned before, Lounesto has presented a compelling argument showcasing an alternative classification framework achieved through the implementation of bilinear covariants \cite{lounestolivro}. Notably, this classification scheme solely relies upon the multi-vector structure inherent in the space-time Clifford algebra. Consequently, the widely acknowledged Lounesto classification offers a highly valuable tool for comprehending and elucidating the properties of spinor fields. Following the usual Lounesto classification, regular spinors are those with at least one of the bilinear covariants $\sigma$ or $\omega$ non-vanishing, encompassing classes 1, 2, and 3. On the other hand, singular spinors are those with both $\sigma$ and $\omega$ vanishing quantities, furnishing the classes 4, 5, and 6.

The six classes of the usual Lounesto classification are all defined below
\begin{enumerate}
  \item $\sigma\neq0$, $\quad \omega\neq0$, \quad $\textbf{J}\neq0, \quad$ $\textbf{K}\neq0,$ $\quad\textbf{S}\neq0$;
  \item $\sigma\neq0$, $\quad \omega=0$, \quad $\textbf{J}\neq0, \quad$ $\textbf{K}\neq0,$ $\quad\textbf{S}\neq0$;
  \item $\sigma=0$, $\quad \omega\neq0$, \quad $\textbf{J}\neq0, \quad$ $\textbf{K}\neq0,$ $\quad\textbf{S}\neq0$;
  \item $\sigma=0=\omega,$ \;\;\;\;\qquad $\textbf{J}\neq0, \quad$  $\textbf{K}\neq0,$ $\quad\textbf{S}\neq0$;
  \item $\sigma=0=\omega,$ \;\;\;\;\qquad $\textbf{J}\neq0, \quad$ $\textbf{K}=0,$ $\quad\textbf{S}\neq0$;
  \item $\sigma=0=\omega,$ \;\;\;\;\qquad $\textbf{J}\neq0, \quad$ $\textbf{K}\neq0,$ $\quad\textbf{S}=0$,
\end{enumerate}
In the context of the Dirac theory, it is pertinent to recall that the aforementioned bilinear covariants are commonly attributed with the standard physical interpretations. However, it is crucial to emphasize that such interpretations are solely applicable to the electron and cannot be generalized beyond that specific context. For complementary material, we refer the reader to the following references \cite{spinorrepresentation,lucaaaca, lucapolar, dale}; they provide some relevant geometric insights into the FPK identities, offering additional perspectives and applications beyond those usually presented in the usual literature.

A fundamental requirement for Lounesto's spinors classification is that all the bilinear covariants must satisfy quadratic algebraic relations, the FPK identities, which read
\begin{eqnarray}
\label{fpk1}
\mathbf{J}^{2}=\omega^{2}+\sigma^{2},\quad\mathbf{K}^{2}=-\mathbf{J}%
^{2},\quad\mathbf{J}\lrcorner\mathbf{K}=0,\quad\mathbf{J}\wedge\mathbf{K}%
=-(\omega+\sigma\gamma_{0123})\mathbf{S}.  
\end{eqnarray}
The three last relations above can be re-written, in terms of the vector components, as follows
\begin{eqnarray}\label{fpkresumida}
K_{\mu}K^{\mu}=-J_{\mu}J^{\mu}, \quad  J_{\mu}K^{\mu} = 0, \quad J_{\mu}K_{\nu}-K_{\mu}J_{\nu} = -\omega S_{\mu\nu} - \frac{\sigma}{2}\epsilon_{\mu\nu\alpha\beta}S^{\alpha\beta}.
\end{eqnarray}
In fact, from the bilinear covariants, we can define a multi-vector object denoted by $\mathbf{Z}$ and known as Fierz aggregate \cite{Crawford:1985qg,lounestolivro} assuming that the bilinear covariants $\sigma,\omega,\mathbf{J},\mathbf{S},\mathbf{K}$ satisfy the FPK identities 
\begin{eqnarray}
\mathbf{Z}=\sigma + \mathbf{J} +i\mathbf{S}+i\mathbf{K}\gamma_{0123}+\omega \gamma_{0123}. \label{Z}
\end{eqnarray} 
We highlight that for singular spinors, the FPK identities are, in general, replaced
by the more general conditions\footnote{Useful relations derived from the identities above (assuming $\mathbf{Z}^{-1}$ exists), read
\begin{align}\label{fpk22}
\mathrm{tr}(\mathbf{Z})=4\sigma ,\nonumber\\ 
\mathrm{tr}(\mathbf{Z}\gamma_{\mu})=4J_{\mu},\nonumber\\
 \mathrm{tr}(\mathbf{Z}i\gamma_{\mu\nu})=4S_{\mu\nu},\\
\mathrm{tr}(\mathbf{Z}i\gamma_{0123}\gamma_{\mu})=4K_{\mu},\nonumber\\
 \mathrm{tr}(\mathbf{Z}\gamma_{0123})=-4\omega.\nonumber
\end{align}} \cite{Crawford:1985qg}:
\begin{align}\label{fpk2}
\mathbf{Z}^{2}=4\sigma \mathbf{Z},\nonumber\\ 
\mathbf{Z}\gamma_{\mu}\mathbf{Z}=4J_{\mu}\mathbf{Z},\nonumber\\
 \mathbf{Z}i\gamma_{\mu\nu}\mathbf{Z}=4S_{\mu\nu}\mathbf{Z},\\
\mathbf{Z}i\gamma_{0123}\gamma_{\mu}\mathbf{Z}=4K_{\mu}\mathbf{Z},\nonumber\\
 \mathbf{Z}\gamma_{0123}\mathbf{Z}=-4\omega \mathbf{Z}.\nonumber
\end{align}
The aggregate plays a central role within the Lounesto classification since, in order to complete the classification itself, $\mathbf{Z}$ have to be promoted to a boomerang, satisfying the relation $\mathbf{Z}^{2}=4\sigma\mathbf{Z}$.
This condition is trivially satisfied for the regular spinors case, and $\mathbf{Z}$ is automatically a boomerang. Nonetheless, for singular spinors case, it is not such a straightforward procedure \cite{lounestolivro}. 

\section{Subclasses: Constraints and Consequences}\label{new_subclasses}

In this section, our aim is to elucidate the potential expansion of Lounesto's classification by scrutinizing the new subclasses unveiled by alternative duals. Moving beyond Lounesto's original framework, which was primarily anchored in the Dirac spinor, and thus in the interpretation of the electron, we will embrace a broader classification encompassing not just particles with electric charges but also those devoid of such charges. Our approach hinges on a meticulous examination of the constraints set forth by the FPK identities. It forbids arbitrary sets of bilinear covariants and, thus, arbitrary classes. After reviewing these constraints, we introduce the following set of allowed subclasses\footnote{Henceforth, as we are taking a broader approach and employing a general dual structure, we will use a distinct notation for the bilinear forms, employing the symbol ``$\sim$'' to differentiate them from the bilinear forms defined with the Dirac dual.} as shown in Table \ref{tab_allowed}. It represents the starting point of any theory in which spinors must obey the FKP identities. Table \ref{tab_forbiden}, on the other hand, represents its complementary set, the set of bilinear covariants configurations such that the corresponding spinors are incompatible with the FPK identities. Such subclasses are thus \emph{forbidden} from the algebraic point of view. More precisely, such subclasses can not have $\tilde{\textbf{J}}=0$ and $\tilde{\textbf{K}}\neq0$ or $\tilde{\textbf{J}}\neq0$ and $\tilde{\textbf{K}}=0$, these conditions violate $\tilde{\mathbf{K}}^{2}=-\tilde{\mathbf{J}}^{2}$, and thus bringing consequences to $\tilde{\sigma}$ and $\tilde{\omega}$ bilinears. Just to mention, there are spinors not obeying the FPK equations \cite{amorph, amorph2}, but we shall pursue the objective of not preclude from the guide provided by algebra. We further develop the discussion of algebraic constraints at the end of the present section.

\begin{table}[H]
\centering
\begin{tabular*}{.5\linewidth}{@{\extracolsep{\fill}}c|cccccc}
\toprule
\textbf{Allowed classes} & $\tilde{\boldsymbol{\sigma}}$ & $\tilde{\boldsymbol{\omega}}$ & $\tilde{\boldsymbol{\textbf{J}}}$ & $\tilde{\boldsymbol{\textbf{K}}}$ & $\tilde{\boldsymbol{\textbf{S}}}$ \\
\midrule
\textbf{1.}   & $\neq0$ & $\neq0$ & $\neq0$ & $\neq0$ & $\neq0$ \\
\emph{1.1}   & $\neq0$ & $\neq0$ & $\neq0$ & $\neq0$ & $=0$ \\
\emph{1.2}   & $\neq0$ & $\neq0$ & $\neq0$ & $=0$ & $\neq0$ \\
\emph{1.3}   & $\neq0$ & $\neq0$ & $\neq0$ & $=0$ & $=0$ \\
\emph{1.4}   & $\neq0$ & $\neq0$ & $=0$ & $\neq0$ & $\neq0$ \\
\emph{1.5}   & $\neq0$ & $\neq0$ & $=0$ & $=0$ & $=0$ \\
\emph{1.6}   & $\neq0$ & $\neq0$ & $=0$ & $=0$ & $\neq0$ \\
\emph{1.7}   & $\neq0$ & $\neq0$ & $=0$ & $\neq0$ & $=0$ \\
\textbf{2.}  & $\neq0$ & $=0$ & $\neq0$ & $\neq0$ & $\neq0$ \\
\emph{2.1}   & $\neq0$ & $=0$ & $\neq0$ & $\neq0$ & $=0$ \\
\textbf{3.}  & $=0$ & $\neq0$ & $\neq0$ & $\neq0$ & $\neq0$ \\
\emph{3.1}   & $=0$ & $\neq0$ & $\neq0$ & $\neq0$ & $=0$ \\
\textbf{4.}  & $=0$ & $=0$ & $\neq0$ & $\neq0$ & $\neq0$ \\
\emph{4.1}   & $=0$ & $=0$ & $=0$ & $\neq0$ & $\neq0$ \\
\textbf{5.}  & $=0$ & $=0$ & $\neq0$ & $=0$ & $\neq0$ \\
\emph{5.1}   & $=0$ & $=0$ & $=0$ & $=0$ & $\neq0$ \\
\textbf{6.}  & $=0$ & $=0$ & $\neq0$ & $\neq0$ & $=0$ \\
\emph{6.1}   & $=0$ & $=0$ & $=0$ & $\neq0$ & $=0$ \\
\textbf{7.}  & $=0$ & $=0$ & $\neq0$ & $=0$ & $=0$ \\
\bottomrule
\end{tabular*}
\caption{All the classes allowed according to FPK identities, including the standard Lounesto classes and their new subclasses.}\label{tab_allowed}
\end{table}
\begin{table}[H]
\centering
\begin{tabular*}{.5\linewidth}{@{\extracolsep{\fill}}c|cccccc}
\toprule
\textbf{Forbiden classes} & $\tilde{\boldsymbol{\sigma}}$ & $\tilde{\boldsymbol{\omega}}$ & $\tilde{\boldsymbol{\textbf{J}}}$ & $\tilde{\boldsymbol{\textbf{K}}}$ & $\tilde{\boldsymbol{\textbf{S}}}$ \\
\midrule
\emph{2.2*} & $\neq0$ & $=0$ & $\neq0$ & $=0$ & $\neq0$ \\
\emph{2.3*} & $\neq0$ & $=0$ & $\neq0$ & $=0$ & $=0$ \\
\emph{2.4*} & $\neq0$ & $=0$ & $=0$ & $\neq0$ & $\neq0$ \\
\emph{2.5*} & $\neq0$ & $=0$ & $=0$ & $\neq0$ & $=0$ \\
\emph{2.6*} & $\neq0$ & $=0$ & $=0$ & $=0$ & $\neq0$ \\
\emph{2.7*} & $\neq0$ & $=0$ & $=0$ & $=0$ & $=0$ \\
\emph{3.2*} & $=0$ & $\neq0$ & $\neq0$ & $=0$ & $\neq0$ \\
\emph{3.3*} & $=0$ & $\neq0$ & $\neq0$ & $=0$ & $=0$ \\
\emph{3.4*} & $=0$ & $\neq0$ & $=0$ & $\neq0$ & $\neq0$ \\
\emph{3.5*} & $=0$ & $\neq0$ & $=0$ & $\neq0$ & $=0$ \\
\emph{3.6*} & $=0$ & $\neq0$ & $=0$ & $=0$ & $\neq0$ \\
\emph{3.7*} & $=0$ & $\neq0$ & $=0$ & $=0$ & $=0$ \\
\bottomrule
\end{tabular*}
\caption{All the subclasses forbidden according to FPK identities, indicating to which of the standard Lounesto classes it belongs.}\label{tab_forbiden}
\end{table}

We shall explicitly construct a realization of Table \ref{tab_allowed} by choosing the appropriate $\Delta$ operator for the dual definition. Our starting point for the general case is the multi-vector elements of the space-time algebra, as shown in equation \eqref{cliffordbasis}, as it generates the whole algebra. Nevertheless, individually, they do not carry any physical interpretation. Thus, we choose to combine them, without loss of generality, with the discrete operators\footnote{A spinorial realization of discrete operators may be given by \cite{Ahluwalia:2019etz} $\mathcal{P}=m^{-1}\gamma_\mu p^{\mu}$ (returning $\gamma_0$ in the rest frame), $\mathcal{C}=\gamma_2 K$, where $K$ complex conjugate from the left, and $\mathcal{T}=-\frac{1}{4!}\epsilon_{\alpha\beta\rho\sigma}\gamma^\alpha\gamma^\beta\gamma^\rho\gamma^{\sigma}\mathcal{C}$. Besides, we use
\begin{equation}
\gamma_0=\left(\begin{array}{cccc}
\mathbb{0}_{2\times 2} & \mathbb{1}_{2\times 2}  \\
\mathbb{1}_{2\times 2} & \mathbb{0}_{2\times 2}
\end{array}\right),
\end{equation} and 
\begin{equation}
\vec{\gamma}=\left(\begin{array}{cccc}
\mathbb{0}_{2\times 2} & \vec{\sigma}  \\
-\vec{\sigma} & \mathbb{0}_{2\times 2}
\end{array}\right).
\end{equation}}. In fact, bearing in mind the following relations: $\{\mathcal{C},\mathcal{P}\}=0$, $[\mathcal{C},\mathcal{T}]=0$ and $[\mathcal{P},\mathcal{T}]=0$, we can check all the possible combinations among discrete symmetries and the Dirac gamma matrices, which may compose the $\Delta$ operator. It brings to light the aforementioned dual operators as shown in Table \ref{tab_allowed2} (we are adopting $\gamma_5:=-i\gamma_{0123}=-i\gamma_{0}\gamma_{1}\gamma_{2}\gamma_{3}$). It should be emphasized that we do not restrict our investigation by covariance symmetries, real bilinears, or any other constraints from standard high-energy physics since, as argued in the Introduction, manageable departures from usual physical requirements are being studied in several different and relevant contexts. We focus on fundamental formal constraints since we aim to uncover possible dual candidates for alternative theories.
\begin{table}[H]
\centering
\begin{tabular*}{\linewidth}{@{\extracolsep{\fill}}c|cccccccccccccc}
\toprule
& $\mathbbm{1}$ & $\mathcal{P}$ & $\mathcal{C}$ & $\mathcal{T}$ & $\mathcal{CP}$ & $\mathcal{CT}$ & $\mathcal{PT}$ & $\mathcal{CPT}$ & $\gamma_0$ & $\gamma_i$  & $\gamma_5$ & $\gamma_{0i}$  & $\gamma_{ij}$ & $\gamma_{5i}$ \\
\midrule
\textbf{$\mathbbm{1}$} & $\mathbbm{1}$ &  $\mathcal{P}$ & $\mathcal{C}$ & $\mathcal{T}$ & $\mathcal{CP}$ & $\mathcal{CT}$ & $\mathcal{PT}$ & $\mathcal{CPT}$ & $\gamma_0$ & $\gamma_i$  & $\gamma_5$ & $\gamma_{0i}$  & $\gamma_{ij}$ & $\gamma_{5i}$ \\
%\midrule
\textbf{$\mathcal{P}$} & $\mathcal{P}$ &  $\mathbbm{1}$ & -$\mathcal{CP}$ & $\mathcal{PT}$ & -$\mathcal{C}$ & -$\mathcal{CPT}$ & $\mathcal{T}$ & -$\mathcal{CT}$ & $\mathcal{P}\gamma_0$ & $\mathcal{P}\gamma_i$ &$\mathcal{P}\gamma_5$ & $\mathcal{P}\gamma_{0i}$ & $\mathcal{P}\gamma_{ij}$ & $\mathcal{P}\gamma_{5i}$ \\
%\midrule
\textbf{$\mathcal{C}$} & $\mathcal{C}$ &  $\mathcal{CP}$ & $\mathbbm{1}$ & $\mathcal{CT}$ & $\mathcal{P}$ & $\mathcal{T}$ & $\mathcal{CPT}$ & $\mathcal{PT}$ & $\mathcal{C}\gamma_0$ & $\mathcal{C}\gamma_i$& $\mathcal{C}\gamma_5$ & $\mathcal{C}\gamma_{0i}$ & $\mathcal{C}\gamma_{ij}$ & $\mathcal{C}\gamma_{5i}$ \\
%\midrule
\textbf{$\mathcal{T}$} &  $\mathcal{T}$ & $\mathcal{PT}$ & $\mathcal{CT}$ & -$\mathbbm{1}$ & $\mathcal{CPT}$ & -$\mathcal{C}$ & -$\mathcal{P}$ & -$\mathcal{CP}$ & $\mathcal{T}\gamma_0$ & $\mathcal{T}\gamma_i$& $\mathcal{T}\gamma_5$ & $\mathcal{T}\gamma_{0i}$ & $\mathcal{T}\gamma_{ij}$ & $\mathcal{T}\gamma_{5i}$ \\
%\midrule
\textbf{$\mathcal{CP}$} & $\mathcal{CP}$&  $\mathcal{C}$ & -$\mathcal{P}$ & $\mathcal{CPT}$ & -$\mathbbm{1}$ & -$\mathcal{PT}$ & $\mathcal{CT}$ & -$\mathcal{T}$& $\mathcal{CP}\gamma_0$ &$\mathcal{CP}\gamma_i$ & $\mathcal{CP}\gamma_5$ & $\mathcal{CP}\gamma_{0i}$ & $\mathcal{CP}\gamma_{ij}$ & $\mathcal{CP}\gamma_{5i}$ \\
%\midrule
\textbf{$\mathcal{CT}$} &$\mathcal{CT}$ & $\mathcal{CPT}$ & $\mathcal{T}$ & -$\mathcal{C}$ & $\mathcal{PT}$ & -$\mathbbm{1}$ & -$\mathcal{CP}$ & -$\mathcal{P}$ & $\mathcal{CT}\gamma_0$ &$\mathcal{CT}\gamma_i$& $\mathcal{CT}\gamma_5$ & $\mathcal{CT}\gamma_{0i}$ & $\mathcal{CT}\gamma_{ij}$ & $\mathcal{CT}\gamma_{5i}$ \\
%\midrule
\textbf{$\mathcal{PT}$} &$\mathcal{PT}$ & $\mathcal{T}$ & -$\mathcal{CPT}$ & -$\mathcal{P}$ & -$\mathcal{CT}$ & $\mathcal{CP}$ & -$\mathbbm{1}$ & $\mathcal{C}$ & $\mathcal{PT}\gamma_0$ &$\mathcal{PT}\gamma_i$ & $\mathcal{PT}\gamma_5$ & $\mathcal{PT}\gamma_{0i}$ & $\mathcal{PT}\gamma_{ij}$ & $\mathcal{PT}\gamma_{5i}$ \\
%\midrule
\textbf{$\mathcal{CPT}$} & $\mathcal{CPT}$&  $\mathcal{CT}$ & -$\mathcal{PT}$ & -$\mathcal{CP}$ & -$\mathcal{T}$ & $\mathcal{P}$ & -$\mathcal{C}$ & $\mathbbm{1}$ & $\mathcal{CPT}\gamma_0$ &$\mathcal{CPT}\gamma_i$ & $\mathcal{CPT}\gamma_5$ & $\mathcal{CPT}\gamma_{0i}$ & $\mathcal{CPT}\gamma_{ij}$ & $\mathcal{CPT}\gamma_{5i}$ \\
\bottomrule
\end{tabular*}
\caption{All possible combinations for the $\Delta$ operator, where $\gamma_i\gamma_j$ is abbreviated as $\gamma_{ij}$ with obvious extension to other $\gamma$'s.}\label{tab_allowed2}
\end{table}

Now, our aim is to filter the relevant dual operators $\Delta$ from Table \ref{tab_allowed2}. It is accomplished by taking into account the algebraic constraint, as given in equation \eqref{algebraic_constrain}. From Table \ref{tab_allowed2}, we check all the combinations holding simultaneously the FPK identities and the algebraic constraint, furnishing the results finally summarized in Table \ref{tab_allowed3}. We emphasize that if two or more operators satisfy the algebraic constraint, then the sum of these operators must also satisfy the given constraint. Thus, we can also define $\Delta$ as being a sum of two or more distinct operators and not only their product.
\begin{table}[H]
\centering
\begin{tabular*}{\linewidth}{@{\extracolsep{\fill}}c|cccccccccccccc}
\toprule
& $\mathbbm{1}$ & $\mathcal{P}$ & $\mathcal{C}$ & $\mathcal{T}$ & $\mathcal{CP}$ & $\mathcal{CT}$ & $\mathcal{PT}$ & $\mathcal{CPT}$ & $\gamma_0$ & $\gamma_i$ & $\gamma_5$ & $\gamma_{0i}$ & $\gamma_{ij}$ & $\gamma_{5i}$ \\
\midrule
\textbf{$\mathbbm{1}$} & $\mathbbm{1}$ &  $\mathcal{P}$ & $\mathcal{C}$ & - & - & - & - & - & $\gamma_0$ & $\gamma_1$, $\gamma_2$ &- & -  & - & - \\
%\midrule
\textbf{$\mathcal{P}$} & $\mathcal{P}$ &  $\mathbbm{1}$ &- & - & - & - & - & - & $\mathcal{P}\gamma_0$ & - &- & $\mathcal{P}\gamma_{01}$, $\mathcal{P}\gamma_{02}$ & - & - \\
%\midrule
\textbf{$\mathcal{C}$} & $\mathcal{C}$ &  - & $\mathbbm{1}$ & - & - & - & - & - & - & $\mathcal{C}\gamma_2$& - & - & $\mathcal{C}\gamma_{12}$ & - \\
%\midrule
\textbf{$\mathcal{T}$} &  - & - & - & -$\mathbbm{1}$ & - & - &- & - & - & -& $\mathcal{T}\gamma_5$ & $\mathcal{T}\gamma_{03}$ & - & $\mathcal{T}\gamma_{52}$ \\
%\midrule
\textbf{$\mathcal{CP}$} & -&  - & - & - & -$\mathbbm{1}$ & - & - & -& $\mathcal{CP}\gamma_0$ & - & - & $\mathcal{CP}\gamma_{02}$ & $\mathcal{CP}\gamma_{13}$ & - \\
%\midrule
\textbf{$\mathcal{CT}$} &- & - & - & - & - & -$\mathbbm{1}$ & - & - & - & - & $\mathcal{CT}\gamma_5$ & $\mathcal{CT}\gamma_{05}$ & - & $\mathcal{CT}\gamma_{51}$, $\mathcal{CT}\gamma_{52}$  \\
%\midrule
\textbf{$\mathcal{PT}$} &- & - & - & - & - & - & -$\mathbbm{1}$ & - & - &$\mathcal{PT}\gamma_3$ & - & $\mathcal{PT}\gamma_{02}$ & $\mathcal{PT}\gamma_{13}$ & $\mathcal{PT}\gamma_{50}$ \\
%\midrule
\textbf{$\mathcal{CPT}$} & -&  - &- & - & - & -& - & $\mathbbm{1}$ & - &- & $\mathcal{CPT}\gamma_5$ & - & $\mathcal{CPT}\gamma_{13}$,  $\mathcal{CPT}\gamma_{23}$  & - \\
\bottomrule
\end{tabular*}
\caption{Combinations for a $\Delta$ operator that satisfies the FKP identities and the algebraic constraint \eqref{algebraic_constrain} simultaneously.}\label{tab_allowed3}.
\end{table}

As it is clear from the above constructions, the new subclasses shown in Table \ref{tab_allowed} can only be accessed if we abandon the Dirac's dual structure and instead employ a more involved dual structure such as $\stackrel{\neg}{\psi}=[\Delta\psi]^{\dag}\gamma_0$, in which $\Delta\neq\mathbbm{1}$. The reader is reefed to \cite{beyondlounesto,rodolfotaka,HoffdaSilva:2022ixq} for explicit examples related to the definition of new dual structures. We emphasize that such a fact can be an aggravating factor when attempting to define the spinor through the Inversion theorem \cite{Takahashi:1982bb} since we must also try to define the dual structure that acts upon the spinor in order to bring it to that class. In this scenario, as discussed, we are abstaining from the usual physics carried by bilinear forms since we are outside the scope carried by Lounesto and its respective interpretations. That being said, it is possible to reach well-behaved bilinear forms even if they are complex numbers. It is to be reinforced, however, that physical scenarios can also be accompanied by such generalizations, as shown in the example presented \cite{Ahluwalia:2020miz}, fitting subclass \emph{1.6}.

Regarding the allowed subclasses, as one can see, class 1 is the most comprehensive one, meaning it has the highest number of possible subclasses due to the conditions of both $\tilde{\sigma}\neq0$ and $\tilde{\omega}\neq0$. This fact opens up the possibility of extending the subclasses to $\tilde{\textbf{J}}=0$ since, according to the FPK identities, if $\tilde{\textbf{J}}^2=0$, then $\tilde{\sigma}=\pm i\tilde{\omega}$. The algebraic conditions reads:
\begin{itemize}
\item[I)] If $\tilde{\textbf{J}}=0$ and $\tilde{\textbf{K}}\neq0$ or $\tilde{\textbf{J}}\neq0$ and $\tilde{\textbf{K}}=0$ or $\tilde{\textbf{J}}=0$ and $\tilde{\textbf{K}}=0$, we are restricted to the following relation $2i\tilde{S}_{\mu\nu}=\epsilon_{\mu\nu\alpha\beta}\tilde{S}^{\alpha\beta}$, where $\tilde{\textbf{S}}=\textbf{0}$ or $\tilde{\textbf{S}}\neq\textbf{0}$.
\item[II)] If $\tilde{\textbf{J}}\neq0$, $\tilde{\textbf{K}}\neq0$ and $\tilde{\textbf{S}}=0$ we are restricted to $\tilde{J}_{\mu}\tilde{K}_{\nu} = \tilde{J}_{\nu}\tilde{K}_{\mu}$.
\end{itemize} 

Classes 2 and 3 require more attention than class 1 case. As one can see, the relation $\tilde{\textbf{J}}^2=0$ is no longer valid. In the class 2 case, we would have $\tilde{\textbf{J}}^2=\tilde{\sigma}^2=0$ leading to $\tilde{\sigma}=0$, and for class 3, we would have $\tilde{\textbf{J}}^2=\tilde{\omega}^2=0$ furnishing $\tilde{\omega}=0$, which are two strong contradictions; another relevant condition that arises from the classes mentioned above is the fact that $\tilde{\textbf{K}}=0$ is not allowed. Such a condition contradicts $\tilde{\textbf{J}}^2=-\tilde{\textbf{K}}^2$, once $\tilde{\textbf{J}}\neq0$. Also, for classes 2 and 3: 
\begin{itemize}
\item[III)] If $\tilde{\textbf{S}}=0$, again, it leads to $\tilde{J}_{\mu}\tilde{K}_{\nu} = \tilde{J}_{\nu}\tilde{K}_{\mu}$, as above.
\end{itemize}

The singular sector of Lounesto classification, classes 4, 5, and 6, exhibit a significant restrictive character, as the only possibility for alteration is to set $\tilde{\textbf{J}}=0$, with all the other bilinear forms unchanged. Any other bilinear form modification would result in the mixing of classes. We highlight that all subclasses hold $\tilde{\mathbf{Z}}^2=4\tilde{\sigma} \tilde{\mathbf{Z}}$.

\section{Physical Impact of new duals}

This section deals with the impact of fermionic quantum fields constructed upon different duals. We perform our analysis by studying the basic but essential properties of such quantum fields, illustrating our point with the field propagator and the spin sums.

We start defining the quantum field and its associated adjoint, respectively, as follows
\begin{eqnarray}\label{campoquantico}
\mathfrak{f}(x) = \int \frac{d^3p}{(2\pi)^3} \frac{1}{\sqrt{2mE(\textbf{p})}}\sum_{h} \bigg[c(\p,h)\psi^{S}_{h}(\textbf{p})e^{-ip_{\mu}x^{\mu}} + d^{\dagger}(\p,h)\psi^{A}_{h}(\textbf{p})e^{ip_{\mu}x^{\mu}}\bigg] 
\end{eqnarray} and
\begin{eqnarray}\label{campoquanticodual}
\stackrel{\neg}{\mathfrak{f}}(x) = \int \frac{d^3p}{(2\pi)^3} \frac{1}{\sqrt{2mE(\textbf{p})}}\sum_{h} \bigg[c^{\dag}(\p,h)\stackrel{\neg}{\psi}^{S}_{h}(\textbf{p})e^{ip_{\mu}x^{\mu}} + d(\p,h)\stackrel{\neg}{\psi}^{A}_{h}(\textbf{p})e^{-ip_{\mu}x^{\mu}}\bigg],
\end{eqnarray} where the upped index $S$ stands for particles and $A$ for anti-particles. It is worth stressing the index $h$ above. In the usual case, in which spin projections perform the only internal degrees of freedom, it stands for this current label. However, whenever degeneracy beyond the spin is in order, under an appropriated pairing of creation and annihilation operators labels and spinors, the $h$ index may be accordingly interpreted as a composite degeneracy label \cite{dharamjhep23}.

The creation and annihilation operators shall obey the usual standard fermionic relations
\begin{eqnarray}
\lbrace c(\p,h),c^{\dag}(\p^{\prime},h^{\prime})  \rbrace = (2\pi)^{3} \delta^3(\textbf{p}-\textbf{p}^{\prime})\delta_{hh^{\prime}}, 
\quad  \lbrace c(\p,h),c(\p^{\prime},h^{\prime})\rbrace = 0 =  \lbrace c^{\dag}(\p,h),c^{\dag}(\p^{\prime},h^{\prime})  \rbrace,
\end{eqnarray} with similar identities assumed for $d(\p,h)$ and $d^{\dag}(\p,h)$ operators. Bearing in mind that the Feynman-Dyson propagator is given by a time-ordered product of $\mathfrak{f}(x)$ and $\stackrel{\neg}{\mathfrak{f}}(x)$, i.e.,
\begin{eqnarray}\label{fdpropagator}
i\mathcal{D}(x-x^{\prime})=\langle 0 \vert \mathfrak{f}(x)\stackrel{\neg}{\mathfrak{f}}(x^{\prime}) \vert 0 \rangle  \theta(x^{0}-x^{\prime 0})- \langle 0 \vert \stackrel{\neg}{\mathfrak{f}}(x^{\prime})\mathfrak{f}(x) \vert 0 \rangle \theta(x^{\prime 0}-x^{0}),
\end{eqnarray}
after a straightforward calculation, we get
\begin{eqnarray}\label{fd3}
\mathcal{D}(x-x^{\prime})&=&\int\frac{d^4 p}{(2\pi)^4}\frac{1}{2E(\boldsymbol{p})}\Bigg[\frac{\sum_{h}\psi^{S}_{h}(\boldsymbol{p})\stackrel{\neg}{\psi}^{S}_{h}(\boldsymbol{p})(p_0+\sqrt{p_{j}p^{j}+m^2})}{p^{2}-m^2+i\epsilon}
\nonumber\\
&+& \frac{\sum_{h}\psi^{A}_{h}(-\boldsymbol{p})\stackrel{\neg}{\psi}^{A}_{h}(-\boldsymbol{p})(p_0-\sqrt{p_{j}p^{j}+m^2})}{p^2-m^2+i\epsilon} \Bigg]e^{-ip_{\mu}(x^{\mu}-x^{\prime\mu})},
\end{eqnarray}
where $j=1, 2, 3$. The propagator defined above is a very general structure; it depends on the spinor and the adjoint definition. In a compact form, one can write the propagator as follows 
\begin{align}\label{propcoms}
D_{\textrm{FD}}(x^\prime-x)\propto\int\frac{\text{d}^4 p}{(2 \pi)^4}\,
e^{-i p_\mu(x^{\prime\mu}-x^\mu)}
\frac{\mathcal{S}(\p)}{p_\mu p^\mu -m^2 + i\epsilon},
\end{align}
in which 
\begin{eqnarray}
\mathcal{S}(\p) \stackrel{def}{=} \frac{1}{2E(\boldsymbol{p})}\Bigg[\sum_{h}\psi^{S}_{h}(\boldsymbol{p})\stackrel{\neg}{\psi}^{S}_{h}(\boldsymbol{p})(p_0+\sqrt{p_{j}p^{j}+m^2})+\sum_{h}\psi^{A}_{h}(-\boldsymbol{p})\stackrel{\neg}{\psi}^{A}_{h}(-\boldsymbol{p})(p_0-\sqrt{p_{j}p^{j}+m^2}) \Bigg].
\end{eqnarray} So far, we have yet to make a particularization to a specific dual. To fix ideas, we now analyze the structure of $\mathcal{S}(\p)$ according to $\Delta=\mathcal{CT}$ as an explicit example\footnote{Among all the possibilities evinced in the last section, we have chosen $\Delta=\mathcal{CT}$ for convenience since its implementation is direct and clear.}.

Hence, particularizing the $\Delta$ operator as before, the only left aspect to be fixed in the analysis is dealing with regular or singular spinors. We shall investigate both cases systematically. Starting with regular spinors we have   
\begin{equation}
\sum_h \psi_h(\p)\stackrel{\neg}{\psi}_h(\p) = i(m\gamma_5-\gamma_{\mu}p^{\mu}),
\end{equation} and the presence of $\gamma_5$ calls attention to a non-invariant spin sum under reflections. Moreover, regarding the propagator's core, $\mathcal{S}(\p)$ operator \eqref{propcoms}, we have, for this case, 
\begin{eqnarray}
\mathcal{S}(\p)=\left(
\begin{array}{cccc}
0 & 0 & -iE & 0 \\
0 & 0 & 0 & -iE \\
iE & 0 & 0 & 0 \\
0 & iE & 0 & 0 \\
\end{array}
\right),
\end{eqnarray}
from which a problematic transformation can be read. The next choice, moving forward, is to deal with singular spinors. For this case, however, the spin sum situation is even worse:
\begin{eqnarray}
&&\sum_h \psi_h(\p)\stackrel{\neg}{\psi}_h(\p) = \nonumber
\\\nonumber
&&\resizebox{\textwidth}{!}{$\left(
	\begin{array}{cccc}
	-\frac{p_x}{2}-\frac{i p_y p_z}{2 (m+E )} & \frac{(m-i p_x-p_y+p_z+E ) (m+i p_x+p_y+p_z+E )}{4 (m+E )} & -\frac{i \left(m^2+m (p_z+E )+p_x^2+p_y^2+p_z (p_z+E )\right)}{2 (m+E )} & -\frac{1}{2} i (p_x-i p_y) \\
	-\frac{(m+i p_x-p_y-p_z+E ) (m-i p_x+p_y-p_z+E )}{4 (m+E )} & \frac{1}{2} \left(p_x+\frac{i p_y p_z}{m+E }\right) & \frac{1}{2} (p_y-i p_x) & -\frac{1}{2} i \left(\frac{p_x^2+p_y^2+p_z^2}{m+E }+m-p_z\right) \\
	\frac{1}{2} i \left(\frac{p_x^2+p_y^2+p_z^2}{m+E }+m-p_z\right) & -\frac{1}{2} i (p_x-i p_y) & -\frac{p_x}{2}+\frac{i p_y p_z}{2 (m+E )} & -\frac{(m-i p_x-p_y-p_z+E ) (m+i p_x+p_y-p_z+E )}{4 (m+E )} \\
	\frac{1}{2} (p_y-i p_x) & \frac{i \left(m^2+m (p_z+E )+p_x^2+p_y^2+p_z (p_z+E )\right)}{2 (m+E )} & \frac{1}{2} \left(\frac{p_x^2+i p_x p_y+p_z^2}{m+E }+m+p_z\right) & \frac{1}{2} \left(p_x-\frac{i p_y p_z}{m+E }\right) \\
	\end{array}
	\right)$},
\end{eqnarray} while the $\mathcal{S}(\p)$ operator reads
\begin{eqnarray}
\mathcal{S}(\p) = 
\left(
\begin{array}{cccc}
-p_x & p_z & -iE & 0 \\
p_z & p_x & 0 & -iE \\
iE & 0 & -p_x & p_z \\
0 & iE & p_z & p_x \\
\end{array}
\right),
\end{eqnarray} showing once again an cumbersome behavior. We alluded to drawbacks of this kind in the Introduction (when dealing with different duals), and the approach to their physical impacts should be in order. Notice that, in the case at hand, one has to face a subtle Lorentz violation, not coming from a tailored Lagrangian term explicitly breaking the symmetry but instead from a violation presented in the spin sums. Nevertheless, the formalism points to an important direction for the case in question. In fact, using the very same spinorial construction of the Ref. (\cite{dharamnpb}, Eqs. (7)-(10)) of one of us, it is possible to make contact with eight singular spinors, addressing the possibility of degeneracy beyond spin in the context of an irreducible representation of the Poincar\'e group. Investigating, then, the spin sums for singular spinors presenting degeneracy beyond the spin, and with duals constructed via $\Delta=\mathcal{CT}$, we have simply
\begin{eqnarray}
\sum_h \psi_h(\p)\stackrel{\neg}{\psi}_h(\p) = -i \gamma_{\mu}p^{\mu},
\end{eqnarray} showing an invariant behavior and leading to $\mathcal{S}(\p)=-i\gamma^{\mu}p_\mu$. The impacts of such a propagator in the complete theory must be further investigated\footnote{For instance, within a well-behaved massless limit case, in which $m\rightarrow 0$ smoothly.}. By now, it can be asserted that the only possibility of using new duals consistently is with concomitant consideration of degeneracy beyond the spin.

The case presented to the dual constructed with $\Delta=\mathcal{CT}$ is not isolated. In fact, as a parenthetical remark, had we analyzed chosen $\Delta=\mathcal{T}$, the spin sums for singular spinors without degeneracy beyond the spin would read 
\begin{eqnarray}
&&\sum_h \psi(\p)_h\stackrel{\neg}{\psi}_h(\p)=\nonumber \\
&&{\footnotesize  \left(
	\begin{array}{cccc}
	p_x+\frac{i p_y p_z}{m+\epsilon } & \frac{-m^2-m p_z-m \epsilon -p_x^2+i p_x p_y-p_z^2-p_z \epsilon }{m+\epsilon } & -\frac{i p_y (m+p_z+\epsilon )}{m+\epsilon } & -m-\frac{p_y (p_y+i p_x)}{m+\epsilon } \\
	\frac{p_x^2+i p_x p_y+p_z^2}{m+\epsilon }+m-p_z & -p_x-\frac{i p_y p_z}{m+\epsilon } & m+\frac{p_y (p_y-i p_x)}{m+\epsilon } & -\frac{i p_y (m-p_z+\epsilon )}{m+\epsilon } \\
	\frac{i p_y (m-p_z+\epsilon )}{m+\epsilon } & -m-\frac{p_y (p_y+i p_x)}{m+\epsilon } & -p_x+\frac{i p_y p_z}{m+\epsilon } & \frac{-p_x^2+i p_x p_y-p_z^2}{m+\epsilon }-m+p_z \\
	m+\frac{p_y (p_y-i p_x)}{m+\epsilon } & \frac{i p_y (m+p_z+\epsilon )}{m+\epsilon } & \frac{p_x^2+i p_x p_y+p_z^2}{m+\epsilon }+m+p_z & p_x-\frac{i p_y p_z}{m+\epsilon } \\
	\end{array}
	\right)}.
\end{eqnarray} Imbued with the idea that spin sums are at the heart of a quantum field construction, we present a table below describing the behavior of spin sums for the relevant duals in the previous section.
\begin{table}[H]
	\centering
	\begin{tabular}{c|ccc}
		\hline 
		\;\;\;\;$\Delta$\;\;\;\; & \;\;\;\;Regular spinors\;\;\;\; &\;\;\;\; Singular Spinors\;\;\;\;& \;\;\;\;Singular Spinors (with degeneracy beyond spin) \;\;\;\; \\ 
		\hline
		$\I$ & $\checkmark$ & $\text{\sffamily X}$ & $\checkmark$\\ 
		
		$\mathcal{P}$ &  $\checkmark$ & $\checkmark$&  $\checkmark$\\ 
		
		$\mathcal{C}$ & $\text{\sffamily X}$ & $\text{\sffamily X}$&$\checkmark$ \\ 
		
		$\mathcal{T}$ & $\text{\sffamily X}$ & $\text{\sffamily X}$ &$\checkmark$\\ 
		
		$\mathcal{CP}$ & $\text{\sffamily X}$ & $\text{\sffamily X}$ &$\checkmark$\\ 
		
		$\mathcal{CT}$ & $\checkmark^*$ & $\text{\sffamily X}$ &$\checkmark$\\ 
		
		$\mathcal{PT}$ & $\text{\sffamily X}$ & $\text{\sffamily X}$&$\checkmark^*$ \\ 
		
		$\mathcal{CPT}$ & $\checkmark^*$  & $\text{\sffamily X}$ &$\checkmark^*$\\ 
		\hline 
	\end{tabular} 
	\caption{Spin sums behavior. Asterisk denotes the cases for which regular spinors have spin sums given by $\gamma_{\mu}p^{\mu}\pm m\gamma_5$, while singular spinors (with degeneracy) have $\pm m \gamma_5$.}
\end{table}

\section{Final Remarks}\label{conclusion}

This research explored extra subclasses within the Lounesto classification, specifically when diverse dual spinors are permitted. As illustrated in Tables \ref{tab_allowed} and \ref{tab_allowed3}, while these newly identified subclasses offer interesting mathematical avenues, it is relatively secure to ensure that none of them shall be relevant within the realm of {\it usual} quantum field theory. As far as we know, there is no definitive assurance that these subclasses encompass spinors with coherent dynamics, especially in the regular sector. Nevertheless, stressing this viewpoint again, the generality provided by the mathematical standpoint allows us to present the new classes to the community as potentially relevant in future studies on theories beyond the standard model. They fundamentally resonate with the algebraic bounds set by the FPK identities, underscoring the intricate balance and relationship between physical phenomena and their mathematical descriptions. The presented analysis points out that when dealing with the investigated different dual structures, an {\it ab initio} choice is in order: to deal with a theory violation Lorentz symmetry at the spin sums level (and therefore with a propagating violation to correlation functions), or the necessity of additional quantum labels beyond the spin. While this analysis does not exhaust all the peculiarities of the quantum formulation based on such duals, it is sufficient to clarify our point, i.e., the possibility of keeping track of the new physical possibilities (and their particularities) by pursuing this algebraically inclined program.

In a scenario where physical theories beyond the standard model are in order, it seems a good (and conservative) approach to speculating new possibilities based on a solid mathematical substrate, serving as a starting point for new developments in theoretical physics.

\section*{Acknowledgements}

The authors acknowledge (in memory) Professor Dharam Vir Ahluwalia, whose invaluable contributions to the field continue to inspire and guide our endeavors. RJBR thanks the generous hospitality offered by UNIFAAT and Professor Renato Medina. JMHS thanks to CNPq (grant No. 307641/2022-8) for financial support.

\bibliography{lounesto_subclasses}

\end{document}